\documentstyle[prl,preprint,aps,epsf]{revtex}
\begin{document}

\title{Gross--Ooguri Phase Transition at Zero and Finite Temperature:
Two Circular Wilson Loop Case}

\author{Hungsoo Kim\raisebox{0.8ex}{1}\footnote[1]
{Email:hskim@hep.kyungnam.ac.kr}, 
D. K. Park\raisebox{0.8ex}{1,2,3}\footnote[2]{Email:dkpark@quark.phy.bnl.gov, 
On 
leave from Kyungnam University, Korea},
S. Tamaryan\raisebox{0.8ex}{3,4}\footnote[3]{Email:sayat@moon.yerphi.am},
and
H.J.W. M\"uller--Kirsten\raisebox{0.8ex}{3}\footnote[4]
{Email:mueller1@physik.uni-kl.de}    }
\address{$^1$ Department of Physics, Kyungnam University, Masan, 631-701,
              Korea   \\
         $^2$ Department of Physics, Brookhaven National Laboratory, 
Upton, NY 11973-5000, USA  \\
         $^3$ Department of Physics,
 University of Kaiserslautern, D-67653 Kaiserslautern, Germany  \\
         $^4$ Theory Department, Yerevan Physics Institute, Yerevan--36,
375036, Armenia}

\maketitle

\maketitle
\begin{abstract}
In the context of $AdS/CFT$ correspondence the two Wilson loop
correlator is examined at both zero and finite temperatures. On the
basis of an entirely 
analytical approach we have found
for Nambu--Goto strings the functional relation
$d S_c^{(Reg)} / dL = 2 \pi k$
between Euclidean action
$S_c$  and loop separation $L$ 
 with integration constant $k$,
 which corresponds 
to the analogous formula for point--particles.
 The physical implications of this 
relation are explored in particular for the Gross--Ooguri phase transition at 
finite temperature.
\end{abstract}

\newpage
Recently much attention has been paid to the Wilson loop correlator in 
large--N gauge theory. This interest results 
mainly from the fact that $AdS/CFT$ 
duality\cite{mal98-1}\cite{mal99} makes it more tractable to understand this
highly nontrivial quantum field theory effect through
 a classical description of 
the string configuration in the $AdS$ background. Using this $AdS/CFT$
duality Maldacena\cite{mal98-2}  was able to calculate
for the first time the expectation value of the rectangular
Wilson loop operator at zero temperature and found that the interquark
potential exhibits the Coulomb type behavior
expected from conformal invariance of the gauge theory,
as well as indications of the screening of the 
charge. Furthermore Witten\cite{witt}
has shown that the $AdS/CFT$ duality can be used to explore
finite temperature behavior of gauge theory by compactifying the
$AdS$ Euclidean time on a circle of radius $\propto$ 
1/temperature. 

Maldacena's computational technique has
already been extended to the finite temperature case 
by replacing the $AdS$ metric by a Schwarzschild--$AdS$ 
metric\cite{rey98,brand98}
which implies the same boundary conditions\cite{haw}. The main 
differences of the finite temperature case from the zero temperature one
are (1) the presence of a maximal separation distance between quarks, and
(2) the appearance of a cusp (or bifurcation point) in the plot of 
interquark potential--vs--interquark distance. These
 differences strongly 
suggest that there is a hidden functional relation between physical
quantities, and indeed such a relation
 has been derived explicitly in Ref.\cite{park00}. 

Here we show that there is a similar  relation
in the two Wilson loop case. It is known\cite{gross98} that the 
correlation function of two circular Wilson loops
 implies a phase
transition analogous to that
between the catenoid as minimal
solution of the area connecting  two
concentric circles \cite{gold} and
 the associated discontinuous Goldschmit solution. 
In fact, this Gross--Ooguri(GO) phase transition takes place 
due to the instability of the classical string configuration. The GO phase 
transition at zero temperature has
been examined in more detail in Ref.\cite{zarembo99}
by solving the
equations of motion of the Nambu--Goto string action.

In the following we approach the GO phase transition at zero temperature 
entirely analytically, which enables us to derive a functional 
relation $d S_c^{Reg} / d L = 2 \pi k$
between the regularized Euclidean classical action $S_c^{Reg}$
and the separation $L$ of the Wilson loops, $k$ being
a constant of integration. We will also approach
the GO phase transition at finite temperature and examine the physical 
implications of the functional relation at the finite temperature 
phase transition. Some related finite temperature aspects have been
considered in Ref.\cite{green}

We use the classical Nambu--Goto action for a string world sheet
\begin{equation}
\label{nambu}
S_{NG} = \int d\tau d\sigma \sqrt{det g_{\mu \nu} \partial_{\alpha} X^{\mu}
                             \partial_{\beta} X^{\nu}}.
\end{equation}
where $g_{\mu \nu}$ is the Euclidean $AdS_5\times{\bf S}^5$ metric.
We want to study the theory of 
a type IIB string with coordinates $X^{\mu}(\tau,\sigma)$
which ends on the near extremal D3--brane,
as the dual of an $SU(N)$ gauge theory in the sense
 of Refs.\cite{mal98-1,mal98-2},
 the case of large $N$, large  't Hooft coupling in the gauge theory
corresponding to weak coupling perturbative string theory
in the semiclassical approximation,
and so to supergravity expanded about
an $AdS_5\times{\bf S}^5$ background.  The quantity
under the square root is the induced metric on the
world sheet of the Nambu--Goto string.

The near extremal Euclidean Schwarzschild--$AdS_5$ metric
of a $D3$--brane (with constant dilaton,
and the $S^5$ coordinates eliminated) \cite{horo91} is given
in Poincar${\acute e}$ coordinates by
\begin{equation}
\label{ads1}
ds_E^2 = \left[ U^2 \left( f(U) dt^2 
         + \sum^3_{i=1}dx_i dx_i \right) + \frac{f(U)^{-1}}{U^2} dU^2 \right].
\end{equation} 
Here $t$ is Euclidean time of target space, and
 we choose $R_{ads} = \alpha^{\prime} = 1$ for simplicity, where
$R_{ads}$ and $\alpha^{\prime}$ are the radius of $AdS_5$ and Regge slope
respectively\cite{mal98-1}.
$U$ is the holographic coordinate\cite{mal98-1}.
 The function $f(U)$ is given
 by $f(U) = 1 - U_T^4 / U^4$ where
$U^2_T$ is a parameter proportional to
the energy density above extremality on the brane\cite{mal98-1};
$U_T$ is proportional to the
external temperature $T$ defined by $T=U_T/\pi R_{ads}$
\cite{brand98} which
enters through the periodic identification of 
$t\rightarrow t+1/T$ to make the horizon at $U=U_T$ regular\cite{gross98}.
The world volume coordinates $t, x_i$ of the brane can be 
regarded as coordinates of the dual $3+1$
dimensional gauge theory.

For simplicity again, we choose cylindrical coordinates
$dx_i dx_i = dx^2 + dr^2 + r^2 d\phi^2$, which results in
\begin{equation}
\label{ads2}
ds_E^2 = \frac{1}{z^2} \left[ (1 - U_T^4 z^4) dt^2 + 
        \frac{dz^2}{1 - U_T^4 z^4} + dx^2 + dr^2 + r^2 d\phi^2 \right]
\end{equation} 
where $z \equiv 1 / U$. Thus the $AdS$ boundary is located at $z = 0$. 

For reasons of symmetry we make the following {\it ansatz} for the
minimal surface\cite{zarembo99}
\begin{equation}
X^0=t=0,\; X^1/r=\phi=\sigma,\; X^2 = x(\tau),\; X^3 = r(\tau),\;
X^4 = z(\tau).
\end{equation}
Then (with elements $g_{\mu\nu}$ of eq.(\ref{ads2}) )
the Nambu--Goto action becomes
\begin{equation}
\label{goto}
S_{NG} = 2 \pi \int d\tau \frac{r}{z^2} \sqrt{\cal L_T}
\end{equation}
where 
\begin{equation}
{\cal L_T} = x^{\prime 2} + r^{\prime 2} + \frac{z^{\prime 2}}{1 - U_T^4 z^4}
\end{equation}
and the prime denotes differentiation with respect to $\tau$.
The Euler--Lagrange equations derived from action (\ref{goto}) are
\begin{eqnarray}
\label{e-leq1}
& & \frac{r}{z^2} \frac{x^{\prime}}{\sqrt{\cal L_T}} = k,  \\  \nonumber
& & \frac{d}{d\tau} \left[ \frac{r r^{\prime}}{z^2 \sqrt{\cal L_T}} \right]
- \frac{\sqrt{\cal L_T}}{z^2} = 0,    \\    \nonumber
& & \frac{d}{d\tau} \left[ \frac{r z^{\prime}}{z^2 \sqrt{\cal L_T} 
                                                (1 - U_T^4 z^4)} \right]
    + \frac{2 r \sqrt{\cal L_T}}{z^3} - 
    \frac{2 U_T^4 r z z^{\prime 2}}{\sqrt{\cal L_T} (1 - U_T^4 z^4)^2} = 0,
\end{eqnarray}
where $k$ is the integration constant arising from the equation of motion
for $x$. 

With a gauge choice $\tau = x$ the equations of motion take the form
\begin{eqnarray}
\label{fixing}
r^{\prime 2} + \frac{z^{\prime 2}}{1 - U_T^4 z^4} + 1&=&\frac{r^2}{k^2 z^4},
                                                   \\   \nonumber
r^{\prime \prime} - \frac{r}{k^2 z^4}&=&0,
                                                   \\   \nonumber
\frac{d}{d\tau} \left[ \frac{z^{\prime}}{1 - U_T^4 z^4} \right]
+ \frac{2 r^2}{k^2 z^5} - \frac{2 U_T^4 z^3 z^{\prime 2}}{(1 - U_T^4 z^4)^2}
&=&0.
\end{eqnarray}

We now assume that two circular Wilson loops 
are located at $x = \pm L / 2$ on the AdS boundary and are linked
by propagation of the closed string in the bulk of $AdS_5\times S^5$.
Then we have to solve Eq.(\ref{fixing}) with the boundary conditions
\begin{eqnarray}
\label{boundary}
r(-L/2)&=&r(L/2) = R,    \\   \nonumber
z(-L/2)&=&z(L/2) = \epsilon \approx 0.
\end{eqnarray}  
Here $R$ and $L$ are respectively the radius of the circular Wilson loops and 
the distance between them. We also introduced the positive
infinitesimal constant $\epsilon$ for the regularization of the
minimal surface area later.

We first consider the zero temperature case ($U_T = 0$). This case has
been
analyzed in Ref.\cite{zarembo99} partly analytically. Here we
approach this case entirely analytically using 
various kinds of elliptic functions. This completely analytical approach
enables one to derive a hidden functional relation explicitly.

For $U_T = 0$ Eqs.(\ref{fixing})
become
\begin{eqnarray}
\label{eq0}
r^{\prime 2} + z^{\prime 2} + 1 - \frac{r^2}{k^2 z^4}&=&0, \\ \nonumber
r^{\prime \prime} - \frac{r}{k^2 z^4}&=&0,  \\   \nonumber
z^{\prime \prime} + \frac{2 r^2}{k^2 z^5}&=&0, 
\end{eqnarray}
and after some manipulations can be shown to 
 yield the equation $r^2 + z^2 + x^2 = a^2$, which is solved by
\begin{eqnarray}
r&=&\sqrt{a^2 - x^2} \cos \theta,   \\   \nonumber
z&=&\sqrt{a^2 - x^2} \sin \theta,
\end{eqnarray}
where $a^2 \equiv R^2 + L^2 / 4$, and $\theta$ obeys
\begin{equation}
\label{teq1}
\theta^{\prime} = \pm \frac{a}{a^2 - x^2}
                  \sqrt{\frac{\cos^2 \theta}{k^2 a^2 \sin^4 \theta} - 1}.
\end{equation}
Here we take the upper sign for $x \in [-L/2, 0]$ and the lower sign for
$x \in [0, L/2]$.
The integration over  $\theta$ can be handled by setting $t=\cos^2\theta$.
Then
\begin{eqnarray}
\label{intt}
I(\theta) &\equiv & \int_0^{\theta} d \theta 
    \frac{\sin^2 \theta}{\sqrt{\cos^2 \theta - k^2 a^2 \sin^4 \theta}}
                                                   \\     \nonumber
&=&\frac{1}{2ak}\int^1_t dt \frac{\sqrt{1-t}}{\sqrt{-t(t-\beta_+)(t-\beta_-)}}
\end{eqnarray}
where
\begin{equation}
\beta_{\pm}= \frac{(2 k^2 a^2 + 1) \pm \sqrt{1 + 4 k^2 a^2}}{2 k^2 a^2}.
\label{be}
\end{equation}
Using the integral formula\cite{byrd71}
\begin{equation}
\int_t^b dt \sqrt{\frac{b - t}{(a - t) (t - c) (t - d)}} = 
(a - b) g \left[ \Pi(\phi, \alpha^2, \kappa) - F(\phi, \kappa) \right]
\end{equation}
where $a > b \geq t \geq c > d$, and 
\begin{eqnarray}
\kappa&=& \sqrt{\frac{(b - c) (a - d)}{(a - c) (b - d)}},\;\; 
g = \frac{2}{\sqrt{(a - c) (b - d)}},     \\    \nonumber
\alpha&=&\sqrt{\frac{b - c}{a - c}}, \;\;
\phi = \sin^{-1} \sqrt{\frac{(a - c) (b - t)}{(b - c) (a - t)}}
\end{eqnarray}
and $\Pi$ and $F$ are elliptic integrals of the third and first kinds 
respectively, one can show that
\begin{eqnarray}
\label{solve1}
I(\theta)
&=& \frac{1}{ka}
\frac{\beta_+ - 1}{\sqrt{\beta_+ - \beta_-}} 
\Bigg[ \Pi \left( \sin^{-1} \sqrt{\frac{(\beta_+ - \beta_-)(1 - \cos^2 \theta)}
                                       {(1 - \beta_-)(\beta_+ - \cos^2 \theta)}
                                 }, \frac{1 - \beta_-}{\beta_+ - \beta_-},
                                    \kappa
           \right)   \\   \nonumber 
& & \hspace{2.5cm} - 
        F \left( \sin^{-1} \sqrt{\frac{(\beta_+ - \beta_-)(1 - \cos^2 \theta)}
                                       {(1 - \beta_-)(\beta_+ - \cos^2 \theta)}
                                 }, \kappa
          \right)                               \Bigg]
\end{eqnarray}
with
$$
a=\beta_+, \;\; b = 1, \;\; c= \beta_-, \;\; d=0,
$$
and
\begin{equation}
\label{miss}
\kappa= \sqrt{\frac{\beta_+ (1 - \beta_-)}{\beta_+ - \beta_-}}.
\end{equation}
One may note that for $k\rightarrow 0$ we have
$\beta_-=0$ and so $\kappa\rightarrow 1$.

Using Eq.(\ref{solve1}) one can integrate Eq.(\ref{teq1}) completely.
The final result for $\theta$ assumes the form
\begin{equation}
\label{final}
I(\theta) = \frac{1}{2ka} \ln \frac{(a + \frac{L}{2}) (a \pm x)}
                         {(a - \frac{L}{2}) (a \mp x)},
\end{equation}
where the upper and lower signs correspond again to 
$x \in [-L/2, 0]$ and $x \in [0, L/2]$ respectively.

From Eqs.(\ref{teq1}) and (\ref{final}) one can show easily 
that $\theta(-L / 2) = \theta(L/2) = 0$ and 
$\theta_0 \equiv \theta(0) = \cos^{-1} \sqrt{\beta_-}$, where the latter
is the result of $\theta^{\prime}(0) = 0$. Inserting $x = 0$ in 
Eq.(\ref{final}) and realizing that 
$\sin^{-1} \sqrt{\frac{(\beta_+ - \beta_-) (1 - \cos^2 \theta)} {(1 - \beta_-)
(\beta_+ - \cos^2 \theta)}} \rightarrow \frac{\pi}{2}$ at this point,
 one
can derive
\begin{equation}
\label{dis1}
{\cal F} = \frac{1}{2} \ln \frac{a + \frac{L}{2}}{a - \frac{L}{2}}
= \ln \frac{\sqrt{R^2 + \frac{L^2}{4}} + \frac{L}{2}}{R}
\end{equation}
where
\begin{equation}
\label{dis2}
{\cal F} = \frac{\beta_+ - 1}{\sqrt{\beta_+ - \beta_-}}
        \left[ \Pi \left( \frac{1 - \beta_-}{\beta_+ - \beta_-}, \kappa \right)
               - K(\kappa)      \right]
\end{equation} 
and $\Pi$ and $K$ are complete elliptic integrals.

Hence the $k$--dependence of $L$ is obtained by solving
\begin{equation}
L = (2 \sinh {\cal F}) R
\end{equation}
numerically. 
Correspondingly the equation can be solved for $L$ as 
a function of $k$. Fig. 1 shows this $L$--dependence of $k$ when $R = 1$.
 Fig. 1 exhibits the two--branch structure described in Ref.
\cite{zarembo99}
and hence shows 
that there is a  maximal distance $L_{\ast}$ for the 
existence of the classical catenoid solution. If $ L > L_{\ast}$, the 
classical catenoid solution becomes unstable and hence the physically 
relevant solution in this case becomes two discontinuous one Wilson
loop solutions, which are the so--called Goldschmit
 discontinuous solutions\cite{gross98}. 

The minimal surface is computed by evaluating
the action (\ref{goto}) for the classical solution given
by Eqs.(\ref{eq0}) to (\ref{teq1}). One obtains
\begin{eqnarray}
\label{claction}
S_c &=& 2\pi\int^{L/2}_{-L/2} dx\frac{r}{z^2}\sqrt{1+(r^{\prime})^2
+(z^{\prime})^2}\nonumber\\
&=& 4 \pi \int_{\frac{\epsilon}{R}}^{\theta_0}d\theta
\frac{\cot^2 \theta}{\sqrt{\cos^2 \theta - k^2 a^2 \sin^4 \theta}}\nonumber\\
&=&\frac{2\pi}{ak}\int^{\cos^2\epsilon/R}_{\beta_-}
\frac{dt}{(t-1)}\frac{\sqrt{-t}}
{\sqrt{(\beta_--t)(t-1)(t-\beta_+)}}
\end{eqnarray}
with angular cutoff $\epsilon/R$.
In a way similar to above we use the integral formulas\cite{form}
\begin{eqnarray}
\label{kap}
& & \int_t^b \frac{dt}{t - c}
\sqrt{\frac{a - t}{(b - t) (t - c) (t - d)}} = \frac{a - b}{b - c} g
                                  \int_0^{u_1(t)} nc^2 u du,
                                                 \\    \nonumber
& & \int nc^2 u du = \frac{1}{\kappa^{\prime 2}}
\left[ \kappa^{\prime 2} u - E(\phi, \kappa) + dn u \hspace{.2cm} tn u \right],
\end{eqnarray}
where $a < b \leq t \leq c < d \;\; with \;\; a=0,\; b=\beta_-,\;
 c=1,\; d=\beta_+$, and 
\begin{eqnarray}
g&=& \frac{2}{\sqrt{(a - c) (b - d)}}, \hspace{.2cm}
\kappa^2 \equiv 1 - \kappa^{\prime 2} = \frac{(b - c) (a - d)}{(a- c) (b - d)},
                                                  \\     \nonumber
\phi&=&\sin^{-1} \sqrt{\frac{(a - c) (b - t)}{(b - c) (a - t)}}, \hspace{.2cm} 
u_1(t) = sn^{-1}[\sin \phi]=sn^{-1}\sqrt{\frac{(\beta_--t)}{t(\beta_--1)}}.
\end{eqnarray}
In Eq. (\ref{kap}) the integration is from $u_1(\beta_-)=0$
to $u_1(\cos^2\epsilon/R)\simeq sn^{-1}(1-\frac{\beta_-}{2(\beta_--1)}
\frac{\epsilon^2}{R^2})$.
Then
one can evaluate Eq.(\ref{claction}) which is
\begin{eqnarray}
\label{conaction}
S_c&=&\frac{4 \pi R}{\epsilon} + S_c^{(Reg)},  \\  \nonumber
S_c^{(Reg)}&=& 4 \pi (1 + 4 k^2 a^2)^{\frac{1}{4}}
\left[ (1 - \kappa^2) K(\kappa) - E(\kappa) \right],
\end{eqnarray}
where $\kappa$ is the same as that of Eq.(\ref{miss})
and $E(0,\kappa)=0, E(\pi/2, \kappa)\equiv E(\kappa), E(1)=1$.
The divergent part of $S_c$ results from $dn u \; tn u$ in Eq.(\ref{kap})
and is obtained  by expanding $tn u$ around $u=K$, i.e.
with $tn u = sn u/cn u\simeq -1/(u-K)\kappa^{\prime}$.
Here $u-K$ is obtained in terms 
of $\epsilon/R$ by expanding $sn u$ around $u=K$ which
gives $1-(u-K)^2\kappa^2/2$
and by then comparing with $sn u_1(\cos^2\epsilon/R)$ above. Thus
$tn u|_{u\rightarrow K}=\sqrt{\frac{1-\beta_-}{\beta_-}}\frac{R}{\epsilon}$.
Multiplying by the factors in front of
 the integral one obtains $4\pi R/\epsilon$.

The area of the discontinuous solution
has been calculated in Ref.\cite{bere99} using
the special conformal transformation. Of course, this can also be 
evaluated directly by considering the case
of one circular Wilson loop. Then the 
minimal surface of the discontinuous solution is 
\begin{eqnarray}
\label{discon}
S_{dc}&=&\frac{4 \pi R}{\epsilon} + S_{dc}^{(Reg)},  \\   \nonumber
S_{dc}^{(Reg)}&=& - 4 \pi.
\end{eqnarray}
One should note that for $k=0$ the expression $S_{c}^{(Reg)}$ is
identical
 with $S_{dc}^{(Reg)}$.
This is expected from the first of Eqs.(\ref{e-leq1}) which indicates
that there is no propagation of the string along $x$ when $k = 0$. 

Fig. 2 shows the $L$--dependence of $S_{c}^{(Reg)}$ and $S_{dc}^{(Reg)}$
and demonstrates the phase transition at a critical value
$L=L_{\ast}$.
One should note that $S_{c}^{(Reg)}$ merges  smoothly with $S_{dc}^{(Reg)}$ at
$L = 0$, which indicates again that $S_{c}^{(Reg)} = S_{dc}^{(Reg)}$
 at $k = 0$.
The appearance of the cusp in $S_{c}^{(Reg)}$ at $L = L_{\ast}$ strongly 
suggests that there is a hidden functional relation\cite{park00} in the
case of the two Wilson loop correlator.

In order to derive this relation explicitly we differentiate
$S_{c}^{(Reg)}$ and $L$ with respect to
the elliptic modulus $\kappa$. This is 
straightforwardly achieved using the various derivative formulas 
of elliptic functions\cite{byrd71}. The final relations are simply
\begin{eqnarray}
\label{hidden1}
\frac{d S_{c}^{(Reg)}}{d \kappa}&=& - \frac{4 \pi \kappa}
                    {(2 \kappa^2 - 1)^{\frac{3}{2}}}
      \left[ K(\kappa) - 2 E(\kappa) \right],
                                            \\   \nonumber
\frac{d L}{d \kappa}&=& - \frac{2 a}{\kappa^{\prime} \sqrt{2 \kappa^2 -1}}
      \left[ K(\kappa) - 2 E(\kappa) \right].
\end{eqnarray}
One should note that the coefficients of
the complete elliptic integrals in the brackets
coincide with each other, which thus yields
after some amazing cancellations\begin{equation}
\label{hidden2}
\frac{d S_{c}^{(Reg)}}{d L} = 2 \pi k.
\end{equation}
As observed in Ref.\cite{park00}, this relation has a
close analogy with the
 point--particle formula $d S_E / d P = {\cal E}$, where $S_E$, $P$ and 
${\cal E}$ are Euclidean action, period and energy of the classical
point--particle.
 It is worthwhile noting that Eq.(\ref{hidden2}) and a
condition $S_{c}^{(Reg)}= S_{dc}^{(Reg)}$ at $L = 0$ 
determine completely the $L$--dependence of $S_{c}^{(Reg)}$ from
the  $k$--dependence of
$L$ since Eq.(\ref{hidden2}) says that 
$S_{c}^{(Reg)} = 2 \pi \int k dL$ up to the constant.
We expect a formula of this type to appear
in many models and thus to have some universal validity.

We now turn to the GO phase transition at finite temperature.
In this case it seems impossible to attack Eqs.(\ref{fixing}) analytically.
However, the analysis of the zero temperature case shows
how one can solve Eqs.(\ref{fixing}) numerically. In fact, one can solve 
the second and third of Eqs.(\ref{fixing}) simultaneously
using the first equation as a relation of boundary conditions at $x = 0$.
Solving these coupled differential equations completely determines
$L$ and $R$. If $R$ is fixed by $R_0$, the only step that remains
is to select the solutions which yield $R = R_0$. 

Fig. 3 shows the 
$L$-dependence of $k$ at finite temperature when $R = 1$. 
Fig. 3 indicates that the peak point moves to the upward,
 and the maximum
distance of the Wilson loop $L_{\ast}$ becomes larger when
the temperature
increases. 

Next we consider the minimal surface area in the finite temperature case.
For this quantity the numerical approach is not a useful tool
in view of the
divergent term which arises in the
course of the calculation. It is, in fact, a formidable 
task to regularize the minimal surface in the numerical technique. 
However, assuming that Eq.(\ref{hidden2}) also holds
 in the  finite temperature
case, one can conjecture the $L$--dependence of $S_c^{(Reg)}$ from 
Fig. 3 since Eq.(\ref{hidden2}) tells
us that  $S_c^{(Reg)} = 2 \pi \int k dL$
up to a constant. To fix this constant we need $S_{dc}^{(Reg)}$ of the 
finite temperature case. 

However, even with the numerical technique 
the computation of $S_{dc}^{(Reg)}$ is not an easy problem 
-- again in view
of the divergent term. In order to 
evaluate $S_{dc}^{(Reg)}$ at finite temperatures we observe that
the minimal surface becomes
\begin{equation}
\label{pcs}
S_c = 4 \pi k \left[ r r^{\prime} \mid_{x = \frac{L}{2}} - 
                    \int_0^{\frac{L}{2}} dx r^{\prime 2} \right].
\end{equation}
This is obtained from Eq.(\ref{goto}) and Eq.(\ref{fixing}) and by
performing an appropriate
partial integration. If one examines the behavior of 
$r(x)$ and $z(x)$ as $x \rightarrow L / 2$, it is easy to 
show that the second term in Eq.(\ref{pcs}) is finite. Hence the divergence 
of $S_c$ is contained in the first term of Eq.(\ref{pcs}). 
In the zero temperature case, for example, one can derive the asymptotic
behavior of $r(x)$ and $z(x)$ for $x \approx L/2$:
\begin{eqnarray}
\label{asymptotic}
r&\approx& R - \frac{1}{2R} \left( \frac{3R}{kL} \right)^{\frac{2}{3}}
  y^{\frac{4}{3}} + \frac{y^2}{2R} + \cdots,  \\   \nonumber
z&\approx& \left( \frac{3R}{kL} \right)^{\frac{1}{3}} y^{\frac{2}{3}}
- \frac{3}{5RkL} y^2 + \frac{1}{2R^2} 
\left( \frac{3R}{kL} \right)^{\frac{1}{3}} y^{\frac{8}{3}} + \cdots,
\end{eqnarray}
where $y^2 \equiv L^2 / 4 - x^2$. It is important to note that the
coefficient of $y^2$ in $r$ is not determined by direct expansion but
from the relation $r^2 + z^2 + x^2 = a^2$, which does not have a 
counterpart in the finite temperature case. Hence in the zero temperature case
$r r^{\prime} \mid_{x = \frac{L}{2}}$ becomes
\begin{equation}
\label{rrp}
r r^{\prime} \mid_{x = \frac{L}{2}} = \frac{R}{k \epsilon} - \frac{L}{2}
\end{equation}
which yields same divergence term $4 \pi R / \epsilon$ and 
$S_c^{(Reg)} = - 2 \pi k L - 4 \pi k \int_0^{\frac{L}{2}} dx r^{\prime 2}$.
If one plots this numerically, one can reproduce $S_c^{(Reg)}$ in Fig. 2.
The important point one should note is that the finite term in
$4 \pi k r r^{\prime} \mid_{x = \frac{L}{2}}, {\it i.e.} -2 \pi k L$, 
becomes zero in the limit $k \rightarrow 0$.
Thus the limit $k \rightarrow 0$
of $S_c^{(Reg)}$, which is nothing but $S_{dc}^{(Reg)}$, originates from the 
second term of Eq.(\ref{pcs}). We believe this property is maintained 
in the finite temperature case. Then one can plot the 
temperature--dependence of $S_{dc}^{(Reg)}$ which is shown in Fig. 4.
 Fig. 4
completely determines the shift constant in 
$S_c^{(Reg)} = 2 \pi \int k dL$. Fig. 5 shows $S_c^{(Reg)}$ at various 
temperatures. As explained, the appearance of the cusp in $S_c^{(Reg)}$
indicates the non--monotonic behaviour of the $k$--dependence of $L$
or $L$--dependence of $k$. 
A similar cusp was observed in Ref.\cite{dorn} in a different calculation.
In Fig. 6 we show the temperature--dependence of the critical point
$L_{\ast}$ at which the phase transition occurs.  In the same plot
we also show the temperature--dependence of the point at
which the action has the same value as that of
the discontinuous solution.

In summary, we analyzed above the Gross--Ooguri phase transition at zero and 
finite temperatures. We obtained the functional relation 
$d S_c^{(Reg)} / dL = 2 \pi k$ which is the Nambu--Goto string analogue
of the formula relating Euclidean action
to period and energy of a classical point--particle\cite{us}.
 It is interesting to investigate this 
point--particle analogy in more detail by calculating the spectrum of 
the fluctuation operator. This work is in progress.

\vspace{1cm}

{\bf Acknowledgement}:  D.K.P. acknowledges support by the
Korea Research Foundation Grant (KRF--2000--D00073).

\begin{figure}
\caption{The $L$--dependence of $k$ at zero temperature
showing the two--branch structure for $L$ less than the
maximum value $L_{\ast}$.}
\end{figure}
\vspace{0.4cm}
\begin{figure}
\caption{$L$--dependence of $S_c^{(reg)}$ and $S_{dc}^{(reg)}$ at zero
temperature demonstrating the appearance of the cusp at the critical
value of $L$ which is the
largest value of $L$ in Fig.1.}
\end{figure}
\vspace{0.4cm}
\begin{figure}
\caption{The $L$--dependence of $k$ at finite temperatures
$\propto U_T=1,2,3$.}
\end{figure}
\vspace{0.4cm}
\begin{figure}
\caption{The temperature--dependence of $S_{dc}^{(reg)}$.}
\end{figure}
\vspace{0.4cm}
\begin{figure}
\caption{$L$--dependence of $S_c^{(reg)}$ at finite temperatures.}
\end{figure}
\vspace{0.4cm}
\begin{figure}
\caption{The plot marked $L_{\ast}$ shows the temperature dependence
of the bifurcation point at which the phase transition occurs.
The plot marked $L_{\ast\ast}$ shows the temperature dependence
of the point $L$ at which the action has the same value as 
that of the discontinuous solution.} 
\end{figure}

\newpage
\epsfysize=20cm \epsfbox{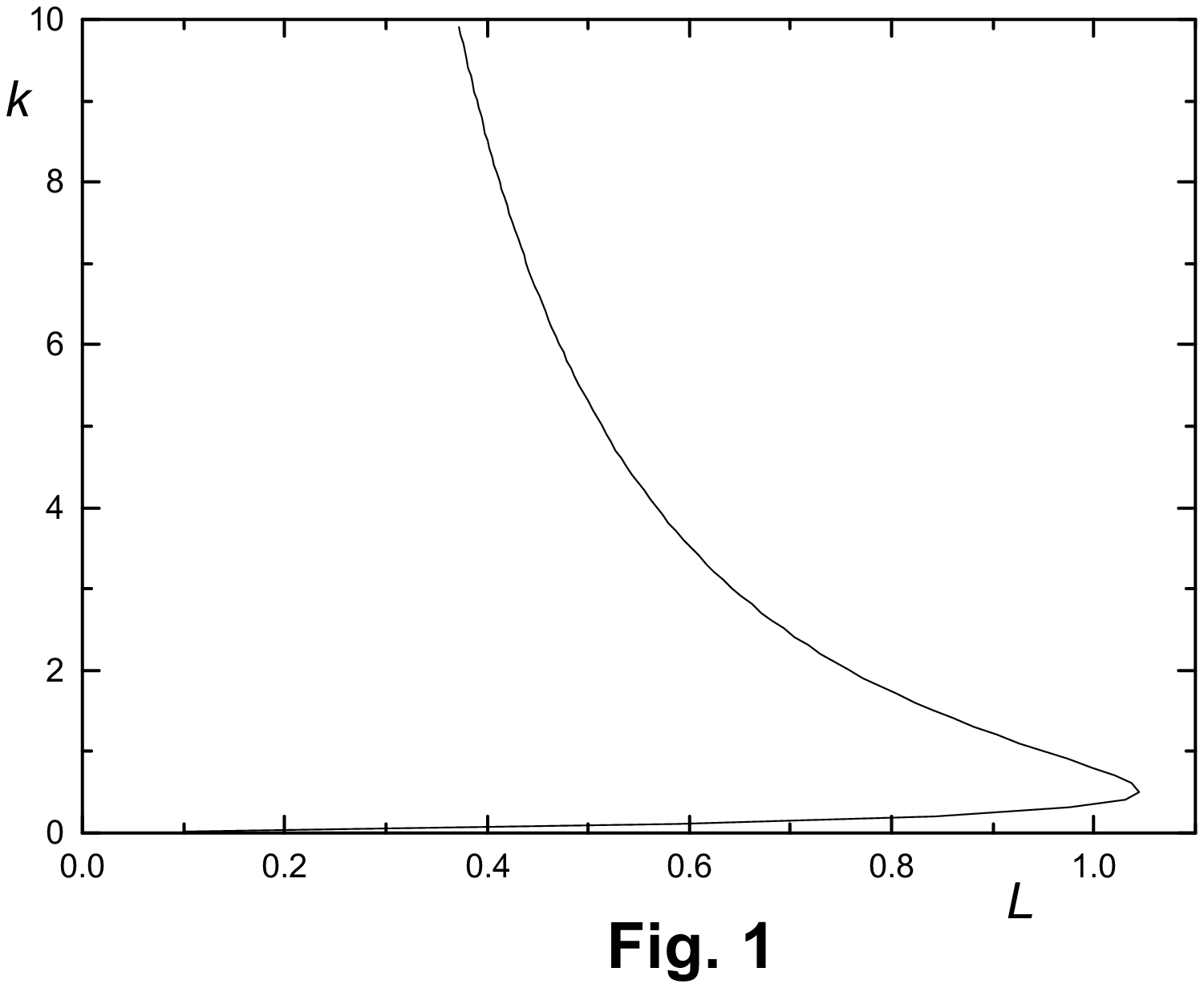}
\newpage
\epsfysize=20cm \epsfbox{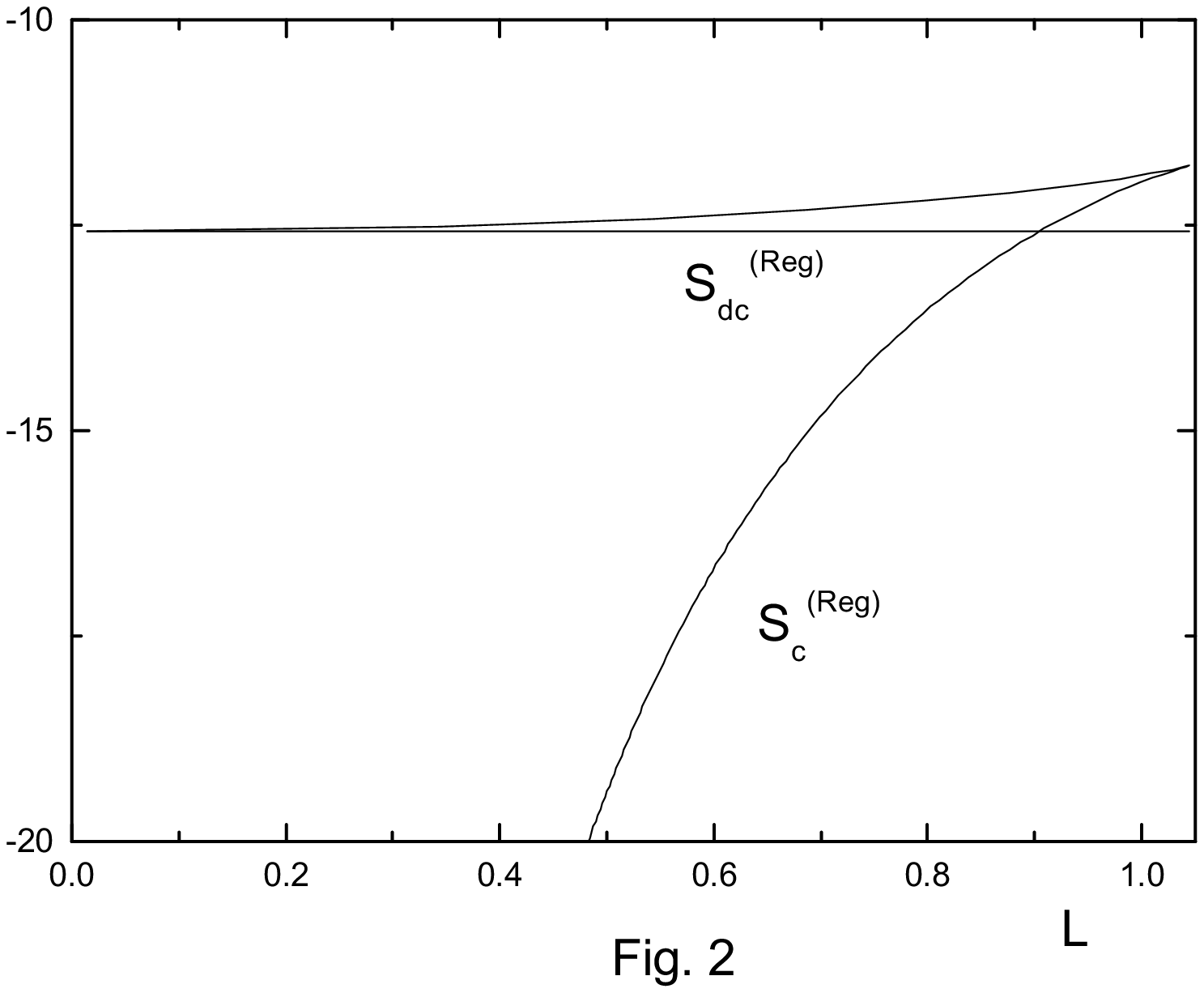}
\newpage
\epsfysize=20cm \epsfbox{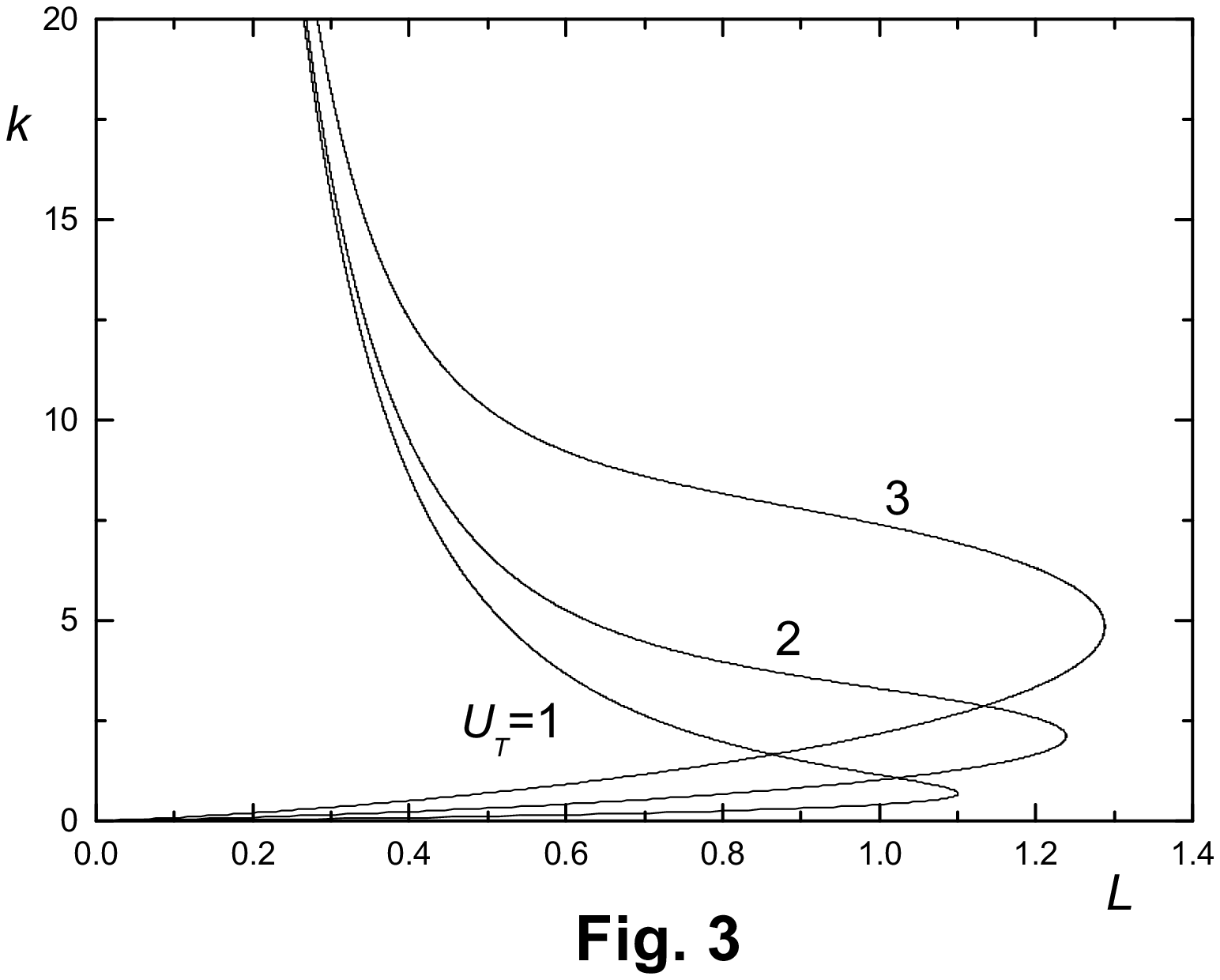}
\newpage
\epsfysize=20cm \epsfbox{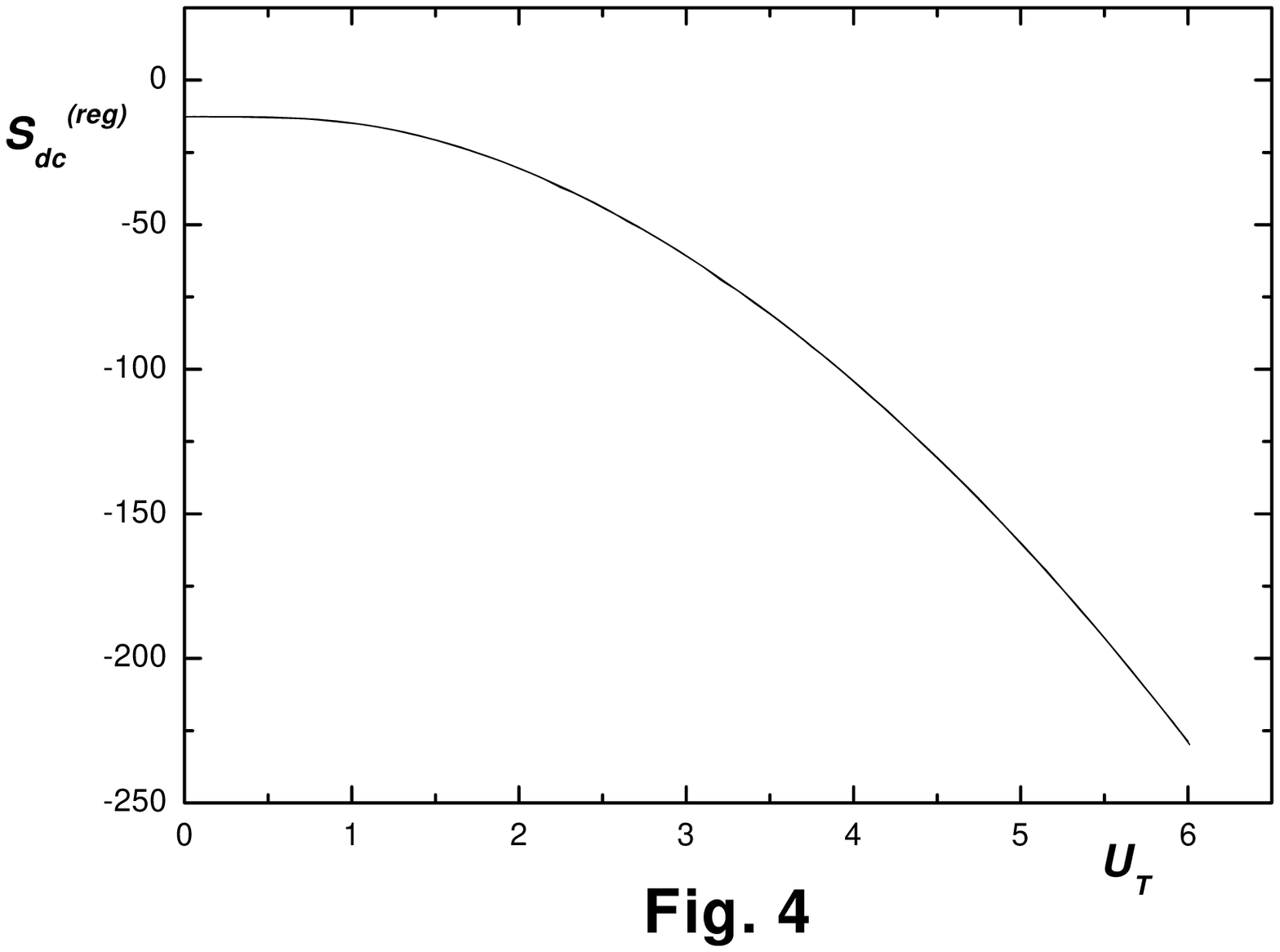}
\newpage
\epsfysize=20cm \epsfbox{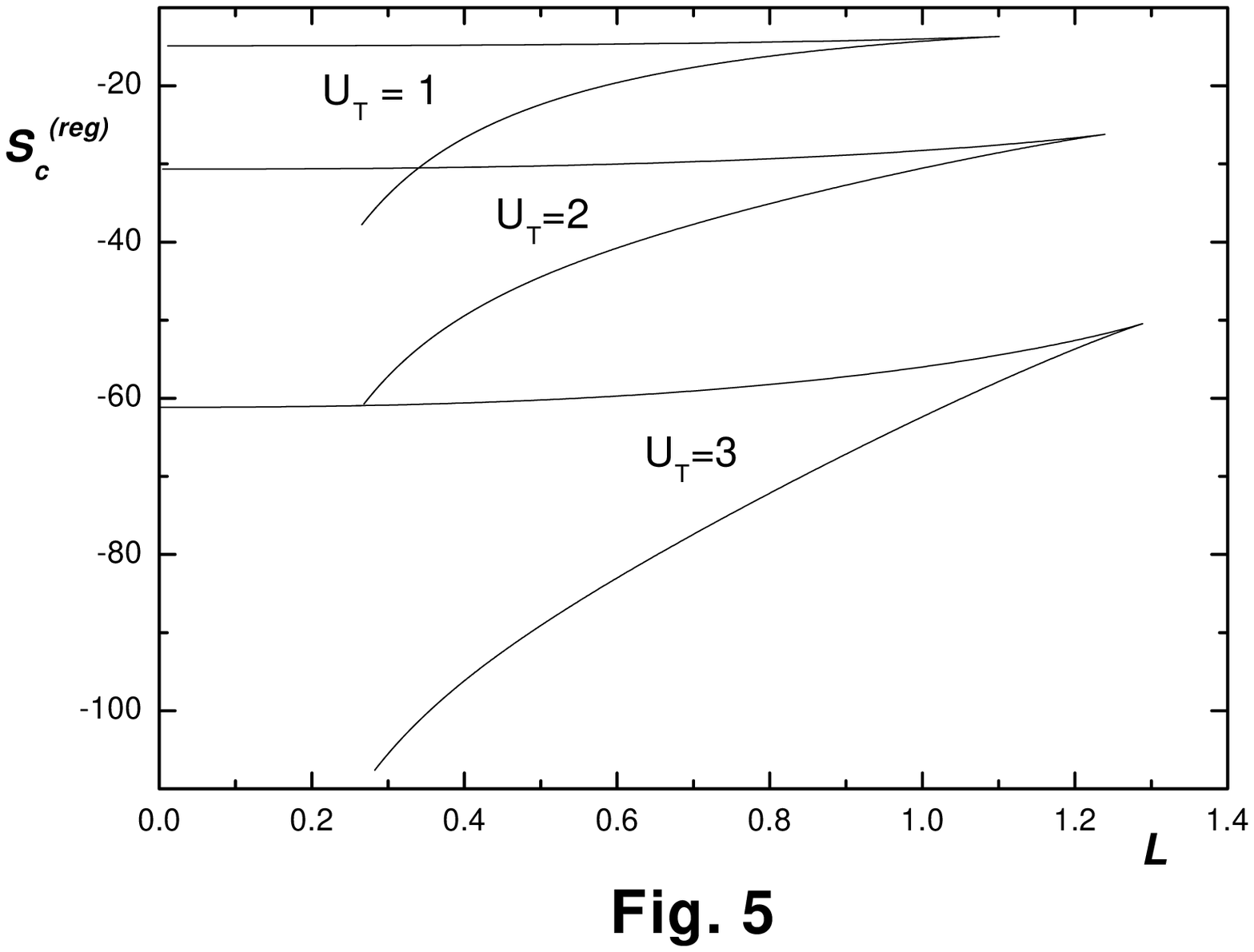}
\newpage
\epsfysize=20cm \epsfbox{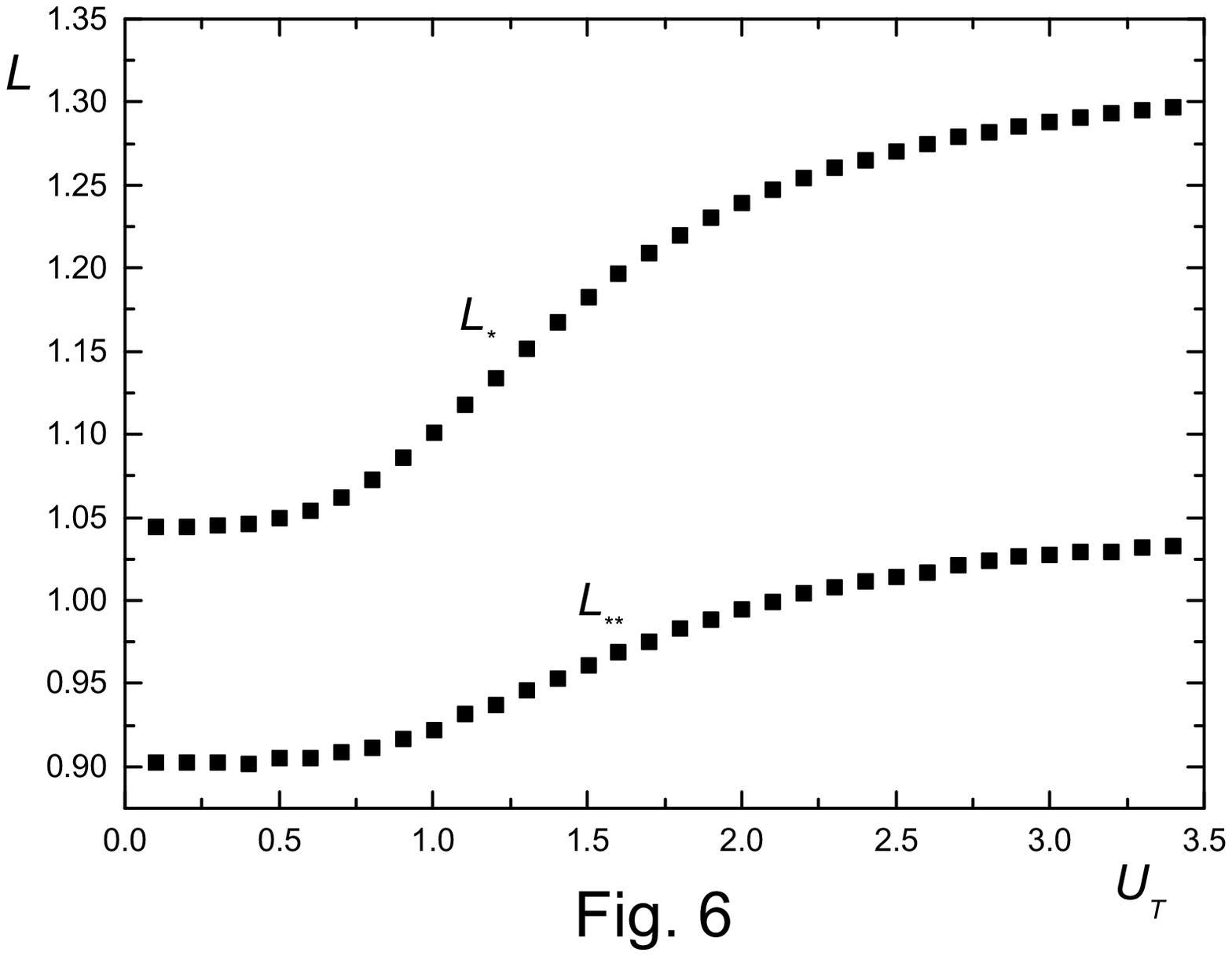}
\end{document}